\documentclass[pra,aps,twocolumn,showpacs]{revtex4}
\usepackage{epsfig,amssymb,amsfonts,amsmath}
\newcommand{\ket}[1]{|#1\rangle} \newcommand{\bra}[1]{\langle #1|}
\newcommand{\proj}[1]{\ket{#1}\bra{#1}}
\begin{document}

\title{Multipartite quantum correlation and entanglement in four-qubit pure states}

\author{Yan-Kui Bai, Dong Yang, and Z. D. Wang}
\email{zwang@hkucc.hku.hk}
 \affiliation{Department of Physics and
Center of Theoretical and Computational Physics, University of
Hong Kong, Pokfulam Road, Hong Kong, China}

\begin{abstract}
Based on the quantitative complementarity relations, we analyze
thoroughly the properties of multipartite quantum correlation and
entanglement in four-qubit pure states. It is found that, unlike
the three-qubit case, the single residual correlation and the
 genuine correlations of three and four qubits  are unable to
quantify entanglement appropriately. More interestingly, from our
qualitative and numerical analysis, it is conjectured that the sum
of all residual correlations is a good quantity for characterizing
the multipartite entanglement in the system.
\end{abstract}


\maketitle

\section{introduction}
Entanglement has been a vital physical resource for quantum
information processing, such as quantum communication
\cite{eke91,ben93} and quantum computation
\cite{ben00,rau01,llb01}. Therefore, the characterization of
entanglement for a given quantum state is a fundamental problem.
Bipartite entanglement is well understood in many aspects
\cite{bdj96,san00,vid02,mbp05}. Especially, for two qubits, its
mixed state entanglement can be characterized with the help of the
so-called concurrence \cite{woo01}. However, in multipartite
cases, the quantification of entanglement is very complicated and
challenging.

A fundamental property of multipartite entangled state is that
entanglement is monogamous. In a three-qubit composite system
$\rho_{ABC}$, the monogamy means that there is a trade-off between
the amount of entanglement that shared by $\rho_{AB}$ and
$\rho_{AC}$, respectively. For the pure state $\ket{\Psi}_{ABC}$,
Coffman, Kundu, and Wootters proved the inequality
$C_{AB}^{2}+C_{AC}^{2}\leq \tau_{A(R_A)}$ \cite{ckw00}, where the
square of the concurrence $C_{ij}$ quantifies the entanglement of
subsystem $\rho_{ij}$ and the linear entropy $\tau_{A(R_A)}$
measures the pure state entanglement between qubit $A$ and
remaining qubits $BC$. Particularly, the residual quantum
correlation in the above equation, \emph{i.e.,} the $3$-tangle
\begin{equation}\label{1}
    \tau(\Psi_{ABC})=\tau_{A(R_A)}-C_{AB}^{2}-C_{AC}^{2},
\end{equation}
was proven to be a good measure for genuine three-qubit entanglement
\cite{ckw00,dur00}. However, in a general case, quantum correlation
and quantum entanglement are inequivalent, although both of them are
nonnegative and invariant under the local unitary (LU)
transformation \cite{ved97,dlz06}. For example, in the Werner state
$\rho_{z}=\frac{1-z}{4}I+z\proj{\psi}$ with
$\ket{\psi}=(\ket{00}+\ket{11})/\sqrt{2}$, the quantum correlation
(quantum discord) \cite{oli02} is greater than $0$ when $z>0$, but
the entanglement (concurrence) is nonzero only when $z>\frac{1}{3}$.
The key difference between the two quantities is that entanglement
does not increase under local operations and classical communication
(LOCC), (i.e., the entanglement monotone property).

Recently, Osborne and Verstraete also proved that the distribution
of bipartite entanglement among $N$-qubit quantum state satisfies
the relation \cite{tjo06}
$C_{A_{1}A_{2}}^{2}+C_{A_{1}A_{3}}^{2}+\cdots+C_{A_{1}A_{N}}^{2}\leq
    \tau_{A_{1}(A_{2}\cdots A_{N})}$, where the
$\tau_{A_{1}(A_{2}\cdots A_{N})}$ is the linear entropy for a pure
state. Comparing with the three-qubit case, it is natural to ask
\emph{whether or not  the residual quantum correlation in an
$N$-qubit pure state ($N>3$) is a good measure of the genuine
multipartite entanglement.}

In this paper, we attempt to answer the above tough question
clearly.  Based on the quantitative complementary relations
(QCRs), we analyze the properties of multipartite correlations and
entanglement in four-qubit pure states. It is shown that the
single residual correlation in the four-qubit case does not
satisfy the entanglement monotone property. In addition, the
genuine three- and four-qubit correlations are unable to quantify
entanglement, either. Finally, in terms of a serious analysis on
the sum of all residual correlations, we conjecture it to be an
appropriate quantity for constituting the multipartite
entanglement measure in the composite system.

The paper is organized as follows. In Sec. II, the properties of
multipartite correlations in four-qubit pure states are analyzed in
detail. As a result, a multipartite entanglement measure is
conjectured. In Sec. III, we give some remarks and main conclusions.
In addition, three examples are given in the Appendix.

\section{Multipartite quantum correlations in four-qubit pure states}

Before analyzing the quantum correlations, we first introduce the
QCRs. Complementarity \cite{boh28} is an essential principle of
quantum mechanics, which is often referred to the mutually
exclusive properties of a single quantum system. As a special
quantum property without classical counterpart, entanglement can
constitute complementarity relations with local properties
\cite{bos02,opp03}. Jakob and Bergou derived a QCR for two-qubit
pure state \cite{jab03}, \emph{i.e.}, $C^{2}+S_k^2=1$, in which
the concurrence $C$ quantifies the non-local correlation of the
two qubits and the $S_{k}^{2}=|\overrightarrow{r_{k}}|^{2}$ is a
measure for single particle characters ($\overrightarrow{r_{k}}$
is the polarization vector of qubit $k$). The experimental
demonstration of this relation was made by Peng \emph{et al}
\cite{pzd05} with nuclear magnetic resonance techniques. For an
$N$-qubit pure state, the generalized QCRs are also available
\cite{pzd05,tes05,cho06}
\begin{eqnarray}\label{2}
  \tau_{k(R_k)}+S_{k}^{2} &=& 1,
\end{eqnarray}
where the linear entropy $\tau_{k(R_k)}=2(1-\mbox{tr}\rho_k^2)$
\cite{san00} characterizes the total quantum correlation between
qubit $k$ and the remaining qubits $R_k$.

For a two-qubit pure state, the linear entropy is a bipartite
quantum correlation. For a three-qubit case, the $\tau_{k(R_k)}$
is composed of the two-qubit and genuine three-qubit correlations
\cite{ckw00}. For an $N$-qubit pure state \cite{czz06}, here we
propose a natural generalization that the linear entropy is
contributed by different levels of quantum correlations,
\emph{i.e.},
\begin{equation}\label{3}
    \tau_{k(R_k)}=t_{N}(\ket{\Psi}_N)+\cdots+\sum_{i< j\in
R_{k}}t_{3}(\rho_{ijk})+\sum_{l\in R_{k}}t_{2}(\rho_{kl}),
\end{equation}
where the $t_{m}$ represents the genuine $m$-qubit quantum
correlation, for $m=2,3,\cdots,N$.
\begin{figure}
\begin{center}
\epsfig{figure=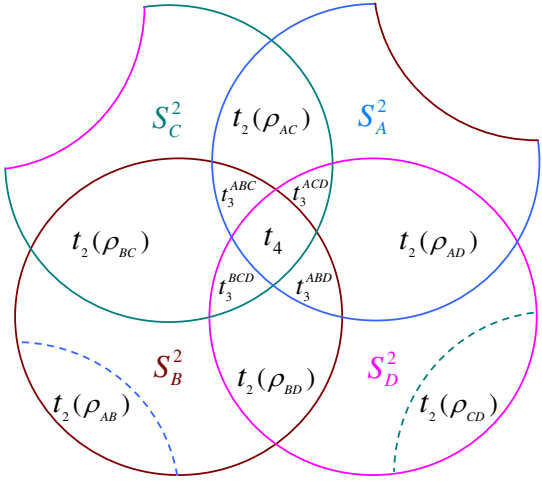,width=0.4\textwidth}
\end{center}
\caption{(Color online) The correlation Venn diagram for a
four-qubit pure state $\ket{\Psi}_{ABCD}$. The overlapping areas
$t_{4}$, $t_{3}$'s, and $t_{2}$'s denote the genuine four-,
three-, and two-qubit quantum correlations, respectively. The
areas without overlapping $S_{k}^{2}$ is the local reality of
qubit $k$, for $k=A,B,C,D$.}
\end{figure}
The Venn diagram, which is often utilized in the set theory, may
be employed to depict quantum correlations in a composite system.
Here we draw schematically a correlation Venn diagram for a
four-qubit pure state $\ket{\Psi}_{ABCD}$ in Fig.1. Qubits $A$,
$B$, $C$, and $D$ are represented by four unit circles,
respectively, and the quantum correlations are denoted by the
overlapping areas of these circles.  According to this diagram,
the four-qubit QCRs can be written as
\begin{eqnarray}\label{4}
t_{4}+t_{3}^{(2)}+t_{3}^{(3)}+t_{3}^{(4)}
+\sum_{l\in R_{A}}t_{2}(\rho_{Al})+S_{A}^2=1,\nonumber\\
t_{4}+t_{3}^{(1)}+t_{3}^{(3)}+t_{3}^{(4)}
+\sum_{l\in R_{B}}t_{2}(\rho_{Bl})+S_{B}^2=1,\nonumber\\
t_{4}+t_{3}^{(1)}+t_{3}^{(2)}+t_{3}^{(4)}
+\sum_{l\in R_{C}}t_{2}(\rho_{Cl})+S_{C}^2=1,\nonumber\\
t_{4}+t_{3}^{(1)}+t_{3}^{(2)}+t_{3}^{(3)} +\sum_{l\in
R_{D}}t_{2}(\rho_{Dl})+S_{D}^2=1,
\end{eqnarray}
where the $t_{3}^{(1)}$, $t_{3}^{(2)}$, $t_{3}^{(3)}$ and
$t_{3}^{(4)}$ are the three-qubit correlations in subsystems
$\rho_{BCD}$, $\rho_{ACD}$, $\rho_{ABD}$, and $\rho_{ABC}$,
respectively. In three-qubit pure states, the quantum correlations
$t_2$ (square of the concurrence) and $t_3$ (3-tangle) in the
linear entropy are good measures for two- and three-qubit
entanglement, respectively. However, it is an open problem that
whether or not the similar relations also hold in a four-qubit
pure state $\ket{\Psi}_{ABCD}$.

Before analyzing the multipartite correlations $t_{4}$ and
$t_{3}^{(i)}$s, we need consider how to evaluate the two-qubit
correlation $t_{2}(\rho_{ij})$ in the pure state
$\ket{\Psi}_{ABCD}$. Similar to the three-qubit case, we make use
of the square of the concurrence which is defined as
$C_{ij}=\mbox{max}[(\sqrt{\lambda_{1}}-\sqrt{\lambda_{2}}-
\sqrt{\lambda_{3}}-\sqrt{\lambda_{4}}), 0]$, where the decreasing
positive real numbers $\lambda_{i}$s are the eigenvalues of matrix
$\rho_{ij}(\sigma_y\otimes\sigma_y)\rho_{ij}^{\ast}(\sigma_y\otimes\sigma_y)$
\cite{woo01}. The main reason for this evaluation is because that
the relation $ \sum_{l\in R_{k}}C_{kl}^{2}=\tau_{k(R_k)}$ holds
for the four-qubit $W$ state
$\ket{\psi}_{ABCD}=\alpha_1\ket{0001}+\alpha_2\ket{0010}+\alpha_3\ket{0100}+\alpha_4\ket{1000}$
which involves only the two-qubit entanglement \cite{ckw00}. In
the following, we will analyze the properties of the single
residual correlation, the genuine three- and four-qubit
correlations, and the sum of all residual correlations,
respectively.

\subsection{Single residual correlation}

Under the above evaluation for the two-qubit quantum correlation,
the multipartite correlation around the qubit $k$ (\emph{i.e.},
the residual correlation) will be
\begin{equation}\label{5}
    M_{k}(\ket{\Psi})=\tau_{k(R_k)}-\sum_{l\in R_{k}}t_{2}(\rho_{kl}),
\end{equation}
in which $t_{2}(\rho_{kl})=C_{kl}^{2}$ and $k=A,B,C,D$. As widely
accepted, a good measure for the multipartite entanglement should
satisfy the following requirements \cite{ved97}: (1) the quantity
should be a non-negative real number; (2) it is unchanged under
the LU operations; (3) it does not increase on average under the
LOCC \emph{i.e.}, the measure is entanglement monotone.

Now we analyze the residual correlation $M_{k}$. According to the
monogamy inequality proven by Osborne and Verstraete \cite{tjo06},
it is obvious that $M_k$ is positive semi-definite. In addition,
for the full separable state and the entangled state involving
only two-qubit correlations, it can be verified that $M_{k}=0$.

The correlation $M_{k}$ is also LU invariant, which can be deduced
from the fact that the linear entropy and the concurrence are
invariant under the LU transformation.

The last condition is that $M_{k}$ should be non-increasing on
average under the LOCC. It is known that any local protocol can be
implemented by a sequence of two-outcome POVMs involving only one
party \cite{dur00}. Without loss of generality, we consider the
local POVM $\{A_{1}, A_{2}\}$ performed on the subsystem $A$,
which satisfies $A_{1}^{\dagger}A_{1}+A_{2}^{\dagger}A_{2}=I$.
According to the singular value decomposition \cite{dur00}, the
POVM operators can be written as $A_{1}=U_{1}diag\{\alpha,
\beta\}V$ and $A_{2}=U_{2}diag\{\sqrt{1-\alpha^2},
\sqrt{1-\beta^2}\}V$, in which $U_{i}$ and $V$ are unitary
matrices. Since $M_{k}$ is LU invariant, we need only  to consider
the diagonal matrices in the following analysis. Note that the
linear entropy and concurrence are invariant under a determinant
one stochastic LOCC (SLOCC) \cite{fer03}, we can deduce
$M_{A}(\ket{\Phi_1})=M_{A}(\frac{A_{1}\ket{\Psi}}{\sqrt{p_{1}}})
=\frac{\alpha^2\beta^2}{p_{1}^{2}}M_{A}(\ket{\Psi})$ and
$M_{A}(\ket{\Phi_2})=M_{A}(\frac{A_{2}\ket{\Psi}}{\sqrt{p_{2}}})
=\frac{(1-\alpha^2)(1-\beta^2)}{p_{2}^{2}}M_{A}(\ket{\Psi})$,
where the $p_{i}=\mbox{tr}[A_{i}\proj{\Psi}A_{i}^{\dagger}]$ is
the normalization factor. After some algebraic deductions similar
to those in Refs. \cite{dur00,won01}, the following relation can
be derived
\begin{equation}\label{6}
    p_{1}M_{A}(\ket{\Phi_{1}})+p_{2}M_{A}(\ket{\Phi_{2}})\leq
    M_{A}(\ket{\Psi}),
\end{equation}
which means the multipartite correlation $M_{A}$ is entanglement
monotone under the local operation performed on subsystem $A$.

It should be pointed out that the above property is \emph{not}
sufficient to show the parameter $M_{A}$ is monotone under the
LOCC. This is because, unlike the three-qubit case, the residual
correlation $M_{k}$ in a four-qubit state will change after
permuting the parties. Therefore, before claiming that the $M_{k}$
is entanglement monotone, one needs to prove the  parameters
$M_{B},M_{C}$, and $M_{D}$ are also non-increasing on average
under the POVM $\{A_{1}, A_{2}\}$ performed on subsystem $A$.
However, this requirement can not be satisfied in a general case,
because the behaviors of the three parameters are quite different
from that of $M_{A}$. For example, in the correlation
$M_{C}=\tau_{C(R_{C})}-C_{AC}^{2}-C_{BC}^{2}-C_{CD}^{2}$, only the
$C_{AC}^{2}$ is invariant under the determinant one stochastic
LOCC performed on subsystem A. With this property, we know
$C_{AC}^{2}$ is entanglement monotone. As to the linear entropy
$\tau_{C(R_C)}$ and the other concurrences ($C_{BC}^{2}$ and
$C_{CD}^{2}$), one can prove that they are decreasing and
increasing under the POVM $\{A_{1},A_{2}\}$, respectively, in
terms of the following two facts: first, for the reduced density
matrices $\rho_{C}$, $\rho_{BC}$ and $\rho_{CD}$, the effect of
the POVM is equivalent to decomposing them into two mixed states,
respectively; second, the linear entropy is concave function and
the concurrence is convex function. Comparing the behaviors of
$M_{A}$ and $M_{C}$ under the POVM, we can not ensure that $M_{C}$
is entanglement monotone (in the Appendix, we give an example in
which the correlation $M_{C}$ will increase under a selected POVM
performed on subsystem $A$). The cases for $M_{B}$ and $M_{D}$ are
similar.

For a kind of symmetric quantum state which has the property
$M_{A}=M_{B}=M_{C}=M_{D}$, is the correlation $M_{k}$ entanglement
monotone? The answer is still negative. Since the symmetry cannot
hold after an arbitrary POVM,  the parameter $M_{k}$ cannot be
guaranteed to be monotone under the next level of POVM once the
property is broken (see such an example in the Appendix).
Therefore, we conclude that the correlation $M_{k}$ is not
entanglement monotone and it is not a good entanglement measure.

\subsection{Three- and four-qubit correlations}

Next, we analyze the properties of the correlations $t_{4}$ and
$t_{3}^{(i)}$. Note that the QCRs  provide only four equations
which cannot determine completely the five multipartite parameters
in general. Therefore, a well-defined measure for $t_3$ or $t_4$
is needed in this case. Recently, an attempt was made to introduce
an information measure $\xi_{1234}$ for the genuine four-qubit
entanglement \cite{czz06}, but this measure can hardly
characterize completely the genuine four-qubit
correlation/entanglement \cite{noteC}.

On the other hand, a mixed $3$-tangle
$\tau_{3}=\mbox{min}\sum_{p_{x},\phi_{x}}p_{x}\tau(\phi_{x})$
\cite{dur00,uhl00} could not be chosen as the correlation $t_{3}$
either, because it is not compatible with the QCRs of Eq.(4). As
an example, we consider the quantum state
$\ket{\psi}_{ABCD}=(\ket{0000}+\ket{1011}+\ket{1101}+\ket{1110})/2$
\cite{fer02}, in which the reduce density matrix $\rho_{BCD}$ can
be decomposed to the mix of two pure states
$\ket{\phi}_{1}=\ket{000}$ and
$\ket{\phi}_{2}=(\ket{011}+\ket{101}+\ket{110})/\sqrt{3}$.
Supposing that the $\tau_{3}$ is a good measure for $t_{3}$, we
can obtain $t_{3}^{(1)}=\tau_{3}(\rho_{BCD})=0$ in terms of the
definition of the mixed $3$-tangle. Then the other multipartite
correlations are determined from Eq. (4), with $t_{4}=1.5$ and
$t_{3}^{(2)}=t_{3}^{(3)}=t_{3}^{(4)}=-0.25$. Because these
correlations are not in the reasonable range, the mixed $3$-tangle
is not a suitable measure compatible with the QCRs.

Although the analytical measures for $t_{4}$ and $t_{3}$ are
unavailable now, we may analyze a special kind of quantum state in
which $t_{4}$ is zero. The quantum state
$\ket{\varphi}=\alpha\ket{0000}+\beta\ket{0101}+\gamma\ket{1000}+\eta\ket{1110}$
is just the case. Suppose that the good correlation measures are
existent and their values correspond to the overlapping regions in
the Venn diagram (Fig.1). It is simple to see that these
correlations are non-negative and LU invariant. In the quantum
state $\ket{\varphi}$, if we let the $t_{3}^{(i)}$ be the
variables, we can obtain the relation
$t_{3}^{(1)}=-\frac{1}{3}t_{4}$ according to the QCRs of Eq. (4).
Due to the non-negative property of the two correlations, we can
judge the four-qubit correlations is zero in this state. Then the
other three-qubit correlations can be solved with the QCRs. In
order to test the entanglement monotone of $t_{3}^{(i)}$ more
clearly, the parameters in $\ket{\varphi}$ are chosen to be
$\alpha=\beta=\gamma=\eta=1/2$ (see the example 3 in the
Appendix). After performing a selected POVM, we find the
$t_{3}^{(2)}$ will increase on average, which implies that  the
correlations $t_{3}$ and $t_{4}$ are not suitable for the
quantification of entanglement.

\subsection{Sum of the residual correlations}

Finally, we consider the sum of all residual correlations,  which
is defined as
\begin{equation}\label{7}
    M=M_{A}+M_{B}+M_{C}+M_{D}=\sum_{k}\tau_{k(R_k)}-2\sum_{p>q}C_{pq}^{2},
\end{equation}
in which $k,p,q=A,B,C,D$. It is obvious that $M$ is nonnegative
and LU invariant in terms of the corresponding properties of
$M_{k}$. It is extremely difficult to prove the entanglement
monotone property analytically. The main hindrance lies in that
one cannot compare the change of the concurrences in a general
quantum state before and after the POVM.

Nevertheless, we conjecture that the correlation $M$ is an
entanglement monotone, as rationalized in some sense below. From
the definition of $M$, it is seen that $M$ is invariant under the
permutations of the subsystems. Without loss of the generality,
suppose that the POVM is performed on the subsystem $A$. In this
case, we analyze the behaviors of the components in $M$. According
to the prior analysis in Eq.(6), the component
$\xi_{1}=\tau_{A(R_A)}-C_{AB}^{2}-C_{AC}^{2}-C_{AD}^{2}$ is
decreasing on average. Moreover, due to the concave property of
linear entropy and the convex property of concurrence, the
component
$\xi_{2}=\tau_{B(R_B)}+\tau_{C(R_C)}+\tau_{D(R_D)}-2(C_{BC}^{2}+C_{BD}^{2}+C_{CD}^{2})$
is also decreasing after the POVM. The only increasing component
is $\xi_{3}=-C_{AB}^{2}-C_{AC}^{2}-C_{AD}^{2}$. It is conjectured
that the decrease of $\xi_{1}$ and $\xi_{2}$ can countervail the
increase of $\xi_{3}$, which results further in the entanglement
monotone property of $M$.

In Fig.2, the quantity $\Delta
M=M(\ket{\Psi})-p_{1}M(\ket{\Phi_{1}})-p_{2}M(\ket{\Phi_{2}})$ is
calculated for nine quantum states
$G_{abcd},L_{abc_{2}},L_{a_{2}b_{2}},L_{ab_{3}},L_{a_{4}},L_{a_{2}0_{3\oplus
1}}, L_{0_{5\oplus 3}},L_{0_{7\oplus 1}}$ and
$L_{0_{3\oplus1}\overline{0}_{3\oplus 1}}$ (the state parameters
we choose are listed in Table I), which are the representative
states under the SLOCC classification (c.f. Ref. \cite{fer02}).
Due to the form of quantum state
$L_{0_{3\oplus1}\overline{0}_{3\oplus 1}}=\ket{0000}+\ket{0111}$,
we perform the POVM on its subsystem $B$. For the other states,
the POVM is performed on the subsystem $A$. From Fig.2, we can see
the correlation $M$ do not increase on average under the POVMs,
which support our conjecture (for the POVMs performed on other
subsystems, we obtain the similar results). In addition, for the
symmetric quantum states $G_{abcd},L_{abc_{2}}$ and $L_{ab_{3}}$,
the second level of the POVM is also calculated and the $\Delta M$
is still nonnegative (in the first level of the POVM performed on
the subsystem $A$, the diagonal elements are $\alpha_{1}=0.4$ and
$\beta_{1}=0.7$; in the second level of POVM, $\alpha_{2}$ and
$\beta_{2}$ are chosen from 0.05 to 0.95, and the interval is
0.01).
\begin{figure}
\begin{center}
\epsfig{figure=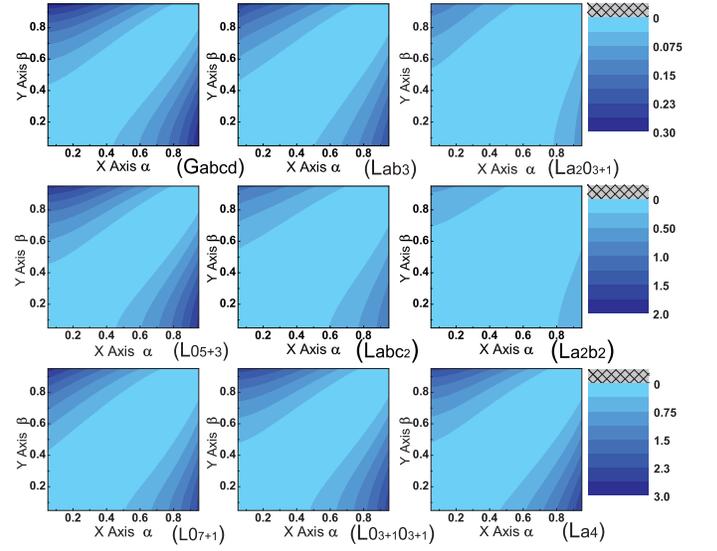,width=0.5\textwidth}
\end{center}
\caption{(Color online) The values of $\Delta M$ for nine
representative states. In the POVM, the diagonal elements $\alpha$
and $\beta$ are chosen from 0.05 to 0.95, and the interval is
0.01. }
\end{figure}
\begin{table}[h]
\begin{center}
\begin{tabular}{|c|c| c| c| c| c|}
\hline\hline $G_{abcd}$ &$L_{abc_{2}}$ & $L_{a_{2}b_{2}}$ &
$L_{ab_{3}}$ & $L_{a_{4}}$ &$L_{a_{2}0_{3\oplus1}}$ \\\hline
$\begin{array}{cc}
  $a=c=1$ \\
  $b=d=0.5$  \\
\end{array}$ & $\begin{array}{cc}
  $a=2$ \\
  $b=c=1$  \\
\end{array}$ & $\begin{array}{cc}
  $a=1$ \\
  $b=1$  \\
\end{array}$ & $\begin{array}{cc}
  $a=1$ \\
  $b=1.5$  \\
\end{array}$ & $a=1$ & $a=1$ \\\hline
\end{tabular}
\caption{The parameters we choose in the quantum states
$G_{abcd},L_{abc_{2}},L_{a_{2}b_{2}},L_{ab_{3}},L_{a_{4}},L_{a_{2}0_{3\oplus
1}}$ (Ref. \cite{fer02}).}
\end{center}
\end{table}

Mainly based on the above analysis, we therefore conjecture that
the multipartite correlation $M$ is entanglement monotone and then
is possible to constitute a measure for the total multipartite
entanglement in four-qubit pure states.

At this stage, we may also introduce the average multipartite
entanglement
\begin{eqnarray}
E_{ms} &=& \frac{M}{4}=\frac{M_{A}+M_{B}+M_{C}+M_{D}}{4},
\end{eqnarray}
to characterize the entanglement per single qubit (ranged in [0,1]),
as far as the correlation $M$ is (conjectured to be) entanglement
monotone. A remarkable merit of this quantity is its computability.
For the quantum state
$G_{abcd}=\frac{a+d}{2}(\ket{0000}+\ket{1111})+\frac{a-d}{2}(\ket{0011}
+\ket{1100})+\frac{b+c}{2}(\ket{0101}+\ket{1010})+\frac{b-c}{2}(\ket{0110}
+\ket{1001})$ which is the generic kind under the SLOCC
classification, the change of $E_{ms}$  with the real parameters $a$
and $d$ are plotted in Fig.3 (the parameters $b=0$ and $c=0.5$ are
fixed). In the regions near ($a=d=0$), ($a\gg c,d$) and ($d\gg
a,c$), the multipartite entanglement $E_{ms}$ tends to zero, which
can be explained that the quantum state tends to be the tensor
product of the two bell states in these ranges. The bigger values of
$E_{ms}$ appear in the regions near ($a=0,d=0.5$), ($a=0.5$ and
$d=0$), and $a=d\gg c$. This is because the quantum state $G_{abcd}$
approaches to the four-qubit $GHZ$ state in these regions( e.g.,
when $a=0$ and $d=0.5$, the $E_{ms}$ is $1$ and the quantum state
can be rewritten as
$G_{abcd}=(\ket{\alpha\alpha\alpha\alpha}+\ket{\beta\beta\beta\beta})/\sqrt{2}$
after a local unitary transformation
$\ket{\alpha}=(\ket{0}+i\ket{1})/\sqrt{2}$ and
$\ket{\beta}=(\ket{0}-i\ket{1})/\sqrt{2}$ ). In this case, the
four-partite entanglement is a dominant one.

\begin{figure}
\begin{center}
\epsfig{figure=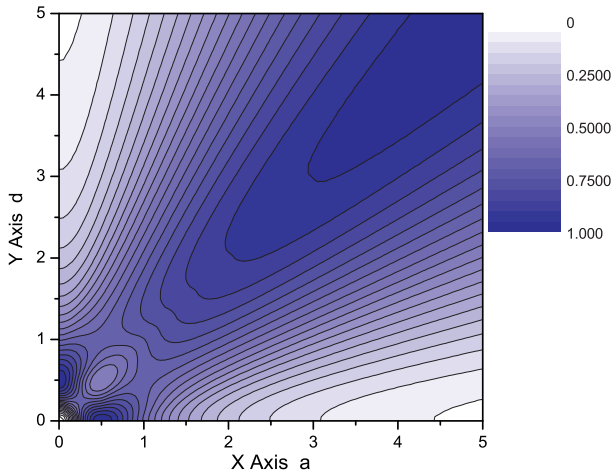,width=0.45\textwidth}
\end{center}
\caption{(Color online) The average multipartite entanglement
$E_{ms}$ for the quantum state $G_{abcd}$ in which the parameters
$a$ and $d$ are chosen from 0 to 5 and the interval is 0.05. The
parameters $b=0$ and $c=0.5$ are fixed.}
\end{figure}

Although the operational meaning of $E_{ms}$ for entanglement
transformation and distillation is not clear now, we can use this
quantity to restrict some procedures which are impossible (suppose
that the $E_{ms}$ is validated to be entanglement monotone). For
example, if the quantity increases in an LOCC transformation from
$\ket{\varphi_{1}}$ to $\ket{\varphi_{2}}$, we can judge that this
procedure is impossible because the entanglement should be
monotone in a real physical transformation.

It should be pointed out that the quantity $E_{ms}$ in Eq. (8)
corresponds to the correlation $t_{4}+\frac{3}{4}\sum
t_{3}^{(i)}$, which is not the total multipartite correlation
$M_{T}=t_{4}+\sum t_{3}^{(i)}$ in the Venn diagram. Whether or not
the $M_{T}$ is a good candidate for the total multipartite
entanglement in the system is worth study in the future. In order
to test the entanglement properties of $M_{T}$, one needs first to
find the appropriate definitions for the correlation $t_{4}$ and
$t_{3}$, respectively.

For an $N$-qubit pure state, the sum of all residual correlations
is given by
\begin{eqnarray}\label{9}
  M_{N}(\Psi_{N}) &=& Nt_{N}+(N-1)\sum t_{N-1}+\cdots+3\sum t_{3}\nonumber \\
  &=& \sum\tau_{k(R_{k})}-2\sum_{i>j} C_{ij}^{2}.
\end{eqnarray}
Similar to the four-qubit case, this quantity is non-negative real
number in terms of the monogamy inequality. In addition, the LU
invariance of $M_{N}$ is guaranteed by the corresponding property
of linear entropy and concurrence. For the entanglement monotone,
we conjecture the correlation $M_{N}$ also satisfies. Therefore,
correlation $M_{N}$ may be able to characterize the multipartite
entanglement in the system. Similarly, the average over $N$ qubits
$M_{N}/N$ (ranged in [0,1]) can be considered as the entanglement
per qubit

\section{discussion and conclusion}

In the correlation Venn diagram of three-qubit pure state
$\ket{\Psi}_{ABC}$ \cite{cho06,cai07},  the quantum correlations at
different levels are able to characterize the corresponding quantum
entanglements. Therefore, the total entanglement in the system is
contributed by the two-qubit entanglement and the genuine
three-qubit entanglement, respectively. However, in the four-qubit
case, the structure of total entanglement is quite complicated; how
to quantify separately the three- and four-qubit entanglement is
still an open problem. It was indicated by Wu and Zhang that the set
of two-, three-, four-partite GHZ states is not a reversible
entanglement generating set for four-party pure states \cite{sjw00}
(\emph{i.e.}, the set of entangled states can not generate an
arbitrary four-party pure state by the LOCC asymptotically
\cite{bpr00}), which implies that the GHZ-class entanglements are
not sufficient for characterizing the structure of total
entanglement in the system. Recently, it was noted by Lohmayer
\emph{et. al.} \cite{loh06} that a kind of rank-2 three-qubit mixed
states which are entangled but do not have the mixed 3-tangle and
concurrence (one can consider that these states are reduced from
four-qubit pure states). This case shows further that the
quantification of entanglement in multi-qubit systems is extremely
complicated and highly nontrivial.

In conclusion, based on the generalized QCRs, we have analyzed the
multipartite correlations in four-qubit pure states. Unlike the
three-qubit case, we find that the similar relations do not hold
again in the four-qubit system. First, the residual correlation
$M_{k}$ is not of entanglement monotone. In addition, the genuine
three- and four-qubit correlations are not suitable to be
entanglement measure, either. Finally, the total residual
correlation $M$ has been analyzed, and it is conjectured that the
average multipartite correlation $E_{ms}$ is able to quantify the
multipartite entanglement in the system.

\section*{Acknowledgments}
The work was supported by the RGC of Hong Kong under grant Nos.
HKU 7051/06P, 7012/06P, and HKU 3/05C, the URC fund of HKU,
NSF-China grants under No. 10429401.

\section*{Appendix}

\textbf{Example 1}: Consider a quantum state
$\ket{\Psi}_{ABCD}=(\ket{0000}+\ket{0011}+\ket{0101}+\ket{0110}
+\ket{1010}+\ket{1111})/\sqrt{6}$, which belongs to the
representative state $L_{a_{2}b_{2}}$ (the parameters is chosen as
$a=b=1$) under the SLOCC classification\cite{fer02}. The POVM
$\{A_{1},A_{2}\}$ is performed on subsystem $A$, which has the
form $A_{1}=U_{1}diag\{\alpha,\beta\}V$ and
$U_{2}diag\{\sqrt{1-\alpha^2},\sqrt{1-\beta^2}\}V$. Due to the LU
invariance of the correlation $M_{k}$, we  need only to consider
the diagonal matrices in which the parameters are chosen to be
$\alpha=0.9$ and $\beta=0.2$. After the POVM, two outcomes
$\ket{\Phi_{1}}=A_{1}\ket{\Psi}/\sqrt{p_{1}}$ and
$\ket{\Phi_{2}}=A_{2}\ket{\Psi}/\sqrt{p_{2}}$ are present, with
the possibilities as $p_{1}=0.5533$ and $p_{2}=0.4467$. Some
calculated results are listed in Table II.
\begin{table}
\begin{center}
\begin{tabular}{c|c c c c c}
\hline\hline $\begin{array}{cc}
   & \mbox{correlation} \\
  \mbox{state} &  \\
\end{array}$ &$\tau_{C(R_C)}$ & $C_{AC}^{2}$ & $C_{BC}^{2}$ &
$C_{CD}^{2}$ &
$M_{C}$ \\\hline $\ket{\Psi}$ & 8/9 & 4/9 & 0 & 0 & 4/9 \\
$\ket{\Phi_{1}}$ & 0.9994 & 0.04703 & 0 & 0 & 0.9524 \\
$\ket{\Phi_{2}}$ & 0.4867 & 0.4063 & 0 & 0 & 0.08042\\ \hline
\end{tabular}
\caption{The values of the correlation measures related to
subsystem $C$ before and after the POVM.} \label{tab1}
\end{center}
\end{table}

According to these values, we can deduce that
$M_{C}(\ket{\Psi})-[p_{1}M_{C}(\ket{\Phi_{1}})+p_{2}M_{C}(\ket{\Phi_{2}})]=-0.1185$,
which means that the correlation $M_C$ is increasing under the
LOCC.

\textbf{Example 2}: Consider a symmetric quantum state
$\ket{\Psi}=(3\ket{0000}+3\ket{1111}-\ket{0011}-\ket{1100}+3\ket{0101}
+3\ket{1010}+\ket{0110}+\ket{1001})/2\sqrt{10}$, which belongs to
the representative state $G_{abcd}$ (the state parameters are
chosen as $a=c=0.5$ and $b=d=1$) \cite{fer02}. According to the
analysis in Sec. II $A$, we know that the correlation $M_{k}$ is
monotone under the first level of the POVM. In this example, we
will show that the correlation $M_{A}$ will be increasing under
the second level of the POVM.

The first level of POVM $\{A_{1},A_{2}\}$ is performed on the
subsystem $A$ in which the diagonal elements are $\alpha=0.3$ and
$\beta=0.8$. After the POVM, two outcomes $\ket{\Phi_{1}}$ and
$\ket{\Phi_{2}}$ can be obtained with the probabilities
$p_{1}=0.3650$ and $p_{2}=0.6350$, respectively. Suppose that
$\ket{\Phi}_{1}$ is gained. Then we do the second level of POVM
$\{A_{11},A_{12}\}$ on the subsystem $C$, in which the diagonal
elements are chosen to be $\alpha_{1}=0.9$ and $\beta_{1}=0.2$.
The outcomes $\ket{\Phi_{11}}$ and $\ket{\Phi_{12}}$ are obtained
with the probabilities $p_{11}=0.1929$ and $p_{12}=0.8071$,
respectively. The calculated results are presented in Table III.

\begin{table}[h]
\begin{center}
\begin{tabular}{c|c c c c c}
\hline\hline $\begin{array}{cc}
   & \mbox{correlation} \\
  \mbox{state} &  \\
\end{array}$ &$\tau_{A(R_A)}$ & $C_{AB}^{2}$ & $C_{AC}^{2}$ &
$C_{AD}^{2}$ & $M_{A}$ \\\hline
$\ket{\Phi_{1}}$ & 0.4324 & 0 & 0.2767 & 0 & 0.1556 \\
$\ket{\Phi_{11}}$ & 0.9960 & 0 & 0.2408 & 0 & 0.7552 \\
$\ket{\Phi_{12}}$ & 0.1565 & 0 & 0.07749 & 0 & 0.07901\\ \hline
\end{tabular}
\caption{The values of the correlation measures related to
subsystem $A$ before and after the second level of the POVM.}
\label{tab3}
\end{center}
\end{table}

Comparing the change of $M_{A}$, we can get
$M_{A}(\ket{\Phi}_{1})-[p_{11}M_{A}(\ket{\Phi_{11}})+p_{12}M_{A}(\ket{\Phi_{12}})]
=-0.05382$. This means that the correlation $M_{A}$ is increasing
under the LOCC, and thus $M_{k}$ is not a good entanglement
measure for the symmetric quantum state.

\textbf{Example 3:} We analyze the quantum state
$\ket{\Psi}_{ABCD}=(\ket{0000}+\ket{0101}+\ket{1000}+\ket{1110})/2$,
which is the representative state $L_{0_{5\oplus 3}}$
\cite{fer02}. The POVM $\{A_{1},A_{2}\}$ is performed on the
subsystem $B$. Due to the LU invariance of the correlations
$t_{4}$ and $t_{3}$, we only consider the diagonal elements of the
operators $A_{1}$ and $A_{2}$ (in the form of the singular value
decomposition) in which the parameters are chosen to be
$\alpha=0.9$ and $\beta=0.4$. After the POVM, two outcomes
$\ket{\Phi_{1}}$ and $\ket{\Phi_{2}}$ are obtained with the
probabilities $p_{1}=0.4850$ and $p_{2}=0.5150$, respectively. In
Table IV, the values of $t_{4}$ and $t_{3}^{(i)}$ for
$\ket{\Psi}$, $\ket{\Phi_{1}}$ and $\ket{\Phi_{2}}$ are listed.
\begin{table}[h]
\begin{center}
\begin{tabular}{c|c c c c c}
\hline\hline $\begin{array}{cc}
   & \mbox{correlation} \\
  \mbox{state} &  \\
\end{array}$ &$t_{4}$ & $t_{3}^{(1)}$ & $t_{3}^{(2)}$ &
$t_{3}^{(3)}$ & $t_{3}^{(4)}$\\\hline
$\ket{\Psi}$ & 0 & 0 & 0.2500 & 0.2500 & 0.2500 \\
$\ket{\Phi_{1}}$ & 0 & 0 & 0.02721 & 0.1377 & 0.1377 \\
$\ket{\Phi_{2}}$ & 0 & 0 & 0.6651 & 0.1504 & 0.1504\\ \hline
\end{tabular}
\caption{The values of the correlation measures $t_{4}$ and
$t_{3}$ before and after the POVM.} \label{tab4}
\end{center}
\end{table}

With these values, we can get
$t_{3}^{(2)}(\ket{\Psi})-[p_{1}t_{3}^{(2)}(\ket{\Phi_{1}})
+p_{2}t_{3}^{(2)}(\ket{\Phi_{2}})]=-0.1057$, which means that the
correlation $t_{3}$ can increase on average under the LOCC and
that it is not a good entanglement measure.


\begin{thebibliography}{99}
\bibitem{noteC}  For instance, when the parameter $a$ is chosen to be $100$
(or to approach the infinite ), the quantum state
                 $\ket{\phi}=a(\ket{0000}+\ket{0101}+\ket{1010}+\ket{1111})+
                 i\ket{0001}+\ket{0110}-i\ket{1011}$ \cite{fer02} tends to be the
                 tensor product of two Bell states, while the measure
                 $\xi_{1234}$ is about $0.99985$ (approaches to $1$) in this state.
\end{thebibliography}
\end{document}